# Role of Information and Communication Technology (ICT) in the Survival of Small and Medium Scale Enterprises (SME's) in Ghana: Evidence from selected Small and Medium Scale Enterprises in New Juaben Municipality, Koforidua

By

Kwaku Nuamah-Gyambrah[1], Martin Offei Otu[2], Florence Agyeiwaa[3]

[1&2] Department of Computer Science, Koforidua Polytechnic, Koforidua, Ghana

[3] All Nations University College, Koforidua, Ghana

Abstract

This study is to examine the role of ICT in the survival of selected SMEs in Koforidua, Ghana The study employed descriptive technique to conduct the survey. Using a sample of 100 SMEs, an accidental sampling of a non-probability technique was used to gathered data and information. The study argues out that majority of the SMEs operators do use at least one ICT tool in supporting their operations within the New Juaben Municipality. The study revealed that ICT is good and helps business survival in difficult times and become competitive in support of literature r

;eviewed. The study suggested that periodic training in the form of workshops and sensitization programs on the benefits and the use ICT resources in business growth strategies should be organized by National Board for Small-Scale Industries (NBSSI). SME operators, can also outsource their ICT delivery systems by engaging ICT consultants in order to avoid the problem of funding relating to the setting up of their own ICT system which usually requires huge initial capital outlay. The primary policy recommendation arising out of this is that applications for SMEs need to be developed using mobile phones.

**Introduction**

It is very obvious that ICT has done great for businesses. In spite of the possible benefits of ICT, presently a debate exists about how their acceptance enhances firm output. Use of and venture in ICT needs balancing investment in skills, organization and innovation and investment and change entails risks and costs as well as bringing potential benefits. By definition, SMEs are businesses that have twenty nine or less number of employees and also posse's plants and other equipment amounting to GH 100,000 as provided in Aryeetey (2005).

Joel and Lussien (2006) further asserted that SMEs are enterprises that are owned on private basis and managed as such. SMEs can be explained by using features of values of assets, sales volume employee number, size of the businesses and other characteristics (Ghana News Agency, 2006). There is no doubt that the 21st century has clinched itself with ICT as one of the driving forces behind accelerated business growth in the world. Laudon and Laudon (2010), explains Information technology (IT) as any form of technology that supports the activities involving the creation, storage, manipulation and communication of information; together with their related methods, management and application. ICT is a fundamental ingredient for globalization. It is against this backdrop that is why the study seeks to investigate the role of ICT on the survival of SME operators in the New Juaben Municipality, Koforidua.

**Statement of problem**

It is very obvious that current businesses do operate and manage their operation using ICT as one of the many tool used to advance the course of business growth. This mean to say that ICT is seen as the engine to speed up the growth of businesses both in the local and international terrain of businesses. It seems SMEs operators often lament of poor profits and high operational cost. Also SMEs in Ghana and for that matter, in New Juaben Municipality primarily employ traditional methodologies and tools to operate. These traditional means of operations very often has resulted in the collapse of these SMEs within the first 2-6 years of existence. Rouf, K. A. (2012), argued that advancing SMEs are the sole way in the attainment of the M DGs and in structuring an international financial approach which rally the wants of the deprived people. Again, most of these SMEs do not integrate ICT tools in their operations because of reasons such as inadequate finances, lack of technical knowledge, seeming lack of applicability of ICT to the business that the SMEs are engaged, rate allusion with ICT and deprived infrastructural growth among others. The study therefore investigates The Role of ICT in the Survival of SMEs in New Juaben Municipality, Koforidua.

**Research Objectives**

The study seeks to achieve the following objectives:

i. To ascertain the level of ICT adoption by SMEs in their business operations
ii. To assess the extent to which the adoption of ICT has led to the achievement of operational excellence among SMEs
iii. To find out the challenges in ICT usage among SMEs

**Research Questions**

The study was guided by the following research questions:

i. What is the level of ICT adoption by SME are their business operations?
ii. To what extent has the adoption of ICT lead to the achievement of operational goals?
iii. What are the challenges of ICT usage among SMEs?

**Literature Review**

**Ghanaian SME Sector**

The experience of SMEs in Ghana dates far into the late 18's to now and has seen continuation over the years. Records show that opportunities abound for the sector in many sections of the economy. They include the area of agriculture, tourism, information technology, services, energy, and manufacturing among others. Ojo (2009) posited that, among the difficulties of development in young nations more clearly, Nigeria seem to sustain the industrial growth schemes. It seems a notion upheld by most of the rising nations and Ghana is no exception. It can be asserted that over the years in operation, Ghana governments have seen to the growth of the private sector of the country and have also give confidence to this same sector to try and help push the country to a ground of relieve to the citizenry. It is also an undeniable fact that the private sector is enormously made up of SMEs. Nevertheless, these SMEs in the sector seem to be overshadowed by varying difficulties in which the SMEs seen to drown. On the other side, also the productive poor and people without government work in Ghana are those actively involved in own job or employment so to improve their livelihood including their relatives as well.

According to Acolatse, (2012) SMEs in recent years have seen a number of demonstrations on the need to increase industrial productivity through a better utilization of the human resources of an organization. This has brought to the forefront different concepts on how this can be affected through instilling in the employees the willingness to work harder and more efficiently. Some have emphasised the human relations aspect whilst others have stressed motivation. SMEs need the supporting of not only the government but also of all in the country to make it sand firm. The assertion upheld by majority of growing nations plus that of Ghana is that the SMEs sector in enraged by series of problems and needs to be attended to with all seriousness and financial support.

Ghana governments have seen to the growth of the private sector of the country and have also give confidence to this same sector to try and help push the country to a ground of relieve to the citizenry. It is also an undeniable fact that the private sector is enormously made up of SMEs. Nevertheless, these SMEs in the sector seem to be overshadowed by varying difficulties in which the SMEs seen to drown. Over two decades, ICT has given culture with a huge compilation of new communication competences. Such includes populace can now talk or exchange verbal words in real-time with many other from dissimilar nations by means of technologies like that an instant messaging, voice over IP (VoIP), and video-conferencing. According to Barman (2010), ICT is a sunshade word that includes any communiqué mechanism, which includes digital radio, digital TV, mobile phones, central processing unit and system hardware and software, satellite systems including a variety of services and software (applications) connected with these elements, such as videoconferencing and electronic ordering system (Barman, 2010).

**Information Communication Technology (ICT) and SME's**

The ability of small and medium-scale business operators to respond rapidly and appropriately to the environmental challenges depends largely on the information systems management to reflect hopes, dreams and realities of real business situation. A substantial portion of business operator's responsibility lies in his creative abilities driven by new knowledge and information. Information communication technology therefore, plays a crucial role in helping small and medium-scale business operators to design and deliver new products and services with unique features and redirecting and redesigning their business processes to meet current changes (Attom, 2008).Small and medium-scale enterprise require information technology infrastructure to provide a solid platform on which their business processes can be built to meet the dynamic business environment they find themselves. Information technology infrastructure can be said to include computer hardware, software, data, storage technology, and networks providing a portfolio of shared IT resources for the enterprise (Attom, 2008).

The advent of modern telecommunication and its associated benefits like faster emails, electronic faxes, social networks etc. time to deliver a service or offer a deliverable or support has been decreased tremendously leading to enhanced customer satisfaction, leading to repeated business and growth and development of firms. Similarly with modern day ICT user of goods and services that are tangible for instance can track the location of their goods at any point in time (Porter, 2005).

According to Acquah (2012), ICT usage in Ghana have undergone through many evolution. In the early 90's ownership of desktop computers were the reserved privileges of the few affluent in the society and bigger firms. Digital phones were almost nonexistent and smart phones not heard of. Only the top class and the middle class could boast of fixed lines popularly called 'land lines' and these were purely used in offices for voice communications and faxes. This restricted the business man or the SME's to doing business based on physical location. Business could not cross markets. Relevant information distributions among SME's were restricted to when the person is available in the office. These led business to not realize optimal profits because certain vital business information was not received on time. These inadvertently affected the growth and development of many SME's. Then came the popularity of desktop computers and mobile phones usage. Desktop computers became so popular in business establishments solely for secretarial purposes namely word processing and spreadsheet management. Many businesses were not utilizing the full benefits of telecommunication together with computers in connectivity until the early 2000. Mobile phones that came in the late nineties were not of much functionality beyond voice communication, short messaging service (SMS) and gaming (Acquah, 2012).

According to Cela (2005), there are more benefits that SMEs may get from ICT. Including the benefits are the following:

i. Improve efficiency and efficacy of operations (Brady et al., 2002).
ii. Enhances the acceptance of original managerial, tactical and decision-making models (Johnston and Lawrence, 2008).
iii. Facilitate the admittance to fresh surroundings including production of fresh operations and business models

**ICT tools for communication among SME's in Ghana**

Including the many benefits or advantages of ICT to businesses around the world, there are also the ideal tools of ICT that businesses do preferred. These ICT tools are many and vary per the kind of businesses that is being carried out. Among the tools include the following:

**Mobile Phones**

It is very obvious that mobile phones are the simplest for ICT that can be employed in the businesses arena around the world of which Ghana is no exception. The benefit are enormous but not without challenges. It is estimated that more that 80% of the small businesses do use mobile phone in their daily operations as well as outs station operations. The dissimilarity is not so spectacular when ordering goods. All the same, there is a disparity with forty eight percent of Small businesses putting to use cell mobile as comparing to thirty six percent using fixed lines (Esselaar et al, 2007).

**Fixed Lines**

Obviously, the cell mobile has become very easily to handle and it role seem very fulfilling the as compared to land line. Fixed line was good and uses to be very important. Adding to the thought that land lines are not in need, is due to the reason that enormous mass do not have direct entrée to land line telephony at any level (Esselaar et al, 2007).

**Internet usage**

Internet proves to be the one kind sophisticated ICT tool that came late but seem to champion the course of SMEs operations than that of the initial tools. It offers divers benefits and also seems very ok to use. However, it is expensive to manage and use as well as compared to mobile phone and fixed line ICT tools. According to them it is very expensive. Regardless of this accession, it is also true that the benefits are great and also helps in making a businesses operate internationally faster than the other means of going to overseas to operate (Esselaar et al, 2007).

**Challenges of Using ICT among SME's**

A challenge identified in the municipality is the general laggard of SMEs in technology adoption. Reasons accounted for this posture is the elevated price of maintaining up to most recent scientific novelty or advancement in this meticulous trade, particularly in the previous decade at what time the speed of scientific progression was extraordinary. According to Curran (1997), one of the reasons for the Small Business Revival was Information Technology and new technologies which provided opportunities for new enterprises to develop and spread. However, the SMEs within the municipality have been battling with the risk of living with outdated technology which has high operational and administrative costs. The investigation showed that these SMEs owners have inadequate managerial experience and for that matter have less knowledge and most often very little interest in the use of technology. (Ricky-Okine, Twum and Owusu, 2014).

Again, the new technology development which is supposed to make the SMEs grow and spread is so sophisticated and complicated that these SMEs operators find it very difficult to apply in their daily operations because of their low levels of education. This challenge has left the SMEs in the municipality dwindling instead of expanding to provide job opportunities to the indigenes as well as generate more revenue for the government.

Braun (2004) identifies limited ICT literacy of SME owners as a challenge that prevents the capacity of SMEs to decide the suitable science-tech and appreciate the real advantages that bring to their business. He further explains that SME businesses are unknown to working a computer. Again they are cynical of the real advantages to its hub trade, and therefore do have the typecast with the intention of ICT is being the sole technology for superior businesses. Inclusive in the challenge is the notion that ICT is kind of demonic and therefore some people do not want to engage that in their businesses at all. This accession is more of religious (Braun, 2004).

Another challenge is the adoption of ICT. While it is an obvious truth that adoption in all areas of businesses seem difficult, the story is not different with regards to ICT. Many often businesses operators do feel good when it comes to changing from one format to another. So it is with ICT. SME operators find it quite uneasy switching from paper work to computing. Knowledge wise, most are not equipped and therefore sees it as a burden to switch. Again, they believe it is also expensive (Braun, 2004).

ICT usage in Ghana has gone through many evolutions. In the early 90's ownership of desktop computers were the reserved privileges of the few affluent in the society and bigger firms. Digital phones were almost nonexistent and smart phones not heard of. Only the top class and the middle class could boast of fixed lines popularly called 'land lines' and these were purely used in offices for voice communications and faxes (Braun, 2004).

Other barriers or problems of AIS faced by the banks include security and trust. Computers in these modern days are now used as instruments of sophisticated white collar crime; McMullan & Perrier (2003). Financial institution loses several millions of dollars to fraudsters who break into the accounting systems of these organizations with relative ease; McMullan & Perrier (2003). Also, organizations whose accountant does not understand how the AIS can be manipulated are defenseless as they have no locks on their doors; Computer Security Institute. 2003

The above barriers have is seen as a huge challenge which affects the capacity of AIS to meet the laid down goals for which they have been designed for. Based on the report from Ghana Banking Survey (2004), specific to banking, these barriers have affected the operations and profit margins of banks in Ghana. Thus, due to the following barriers above, the banking industry in Ghana has not yet been able to exploit AIS to the fullest. Yet another challenge is the new technology development which is supposed to make the SMEs grow and spread is so sophisticated and complicated that these SMEs operators find it very difficult to apply in their daily operations because of their low levels of education. This challenge has left the SMEs in the municipality dwindling instead of expanding to provide job opportunities to the indigenes as well as generate more revenue for the government.

**Research Methodology**

The study employed quantitative and expressive design. Questionnaires were used as the data collection tool to solicit data and information. The study employed expressive design to gain insight into the role of ICT in survival of SMEs. The population for the study included both registered and non registered Small and Medium Scale Enterprises within the New Juaben Municipality using at least one tool of ICT. The population was one hundred (100) SMEs selected in the New Juaben Municipality. Accidental sampling technique, a non probability sampling method was employed in selecting the SMEs. This was because of the scatter nature of the SME's. Out of 100 copies of questionnaire distributed 60 copies were retrieved completely filled. The computer software SPSS version 16.0 was used to analysis the data collected.

**Findings**

**Objective one:** To find out the level of Information And Communication Technologies now adopted by Small and Medium Enterprises in the New Juaben Municipality.

Table 4.1 Adoption of ICT and its usage

|       |       | Frequency | Percent | Valid Percent | Cumulative Percent |
|-------|-------|-----------|---------|---------------|--------------------|
| Valid | Yes   | 53        | 88.3    | 88.3          | 88.3               |
|       | No    | 7         | 11.7    | 11.7          | 100.0              |
|       | Total | 60        | 100.0   | 100.0         |                    |

**Source: field data, October, 2014**

In view of this question, the researcher decided to inquire if respondents have some knowledge about Information And Communication Technology. From the data collected majority said that they have an idea about ICT and this represented 88.3% (53 respondents) of the total sample size. 7 respondents representing 11.7% Yes. The inference drawn here is majority of the SMEs in New Juaben Municipality do know and have an idea of what ICT. It also shows that these SMEs owners are abreast with technology in some aspect of their businesses. As a follow up question, the researcher asked whether the owners do use any ICT tool in their business operation and from the data gathered, 53 (88.3%) respondents said yes. This represents the majority of the sample size. 7(11.7%) respondents said no, meaning they do not use any ICT tool in their operation. The inference drawn from the analysis is that, majority of the SME operators do use at least one ICT tool in supporting their operations within the new Juaben Municipality. This shows that the SMEs owners understand the need to employ ICT in their operations.

**Table 4.3 ICT tool used in business operations**

|       |              | Frequency | Percent | Valid Percent | Cumulative Percent |
|-------|--------------|-----------|---------|---------------|--------------------|
| Valid | Internet     | 9         | 15.0    | 15.0          | 15.0               |
|       | Mobile Phone | 37        | 61.7    | 61.7          | 76.7               |
|       | E-Money Sys. | 8         | 13.3    | 13.3          | 90.0               |
|       | Fixed line   | 6         | 10.0    | 10.0          | 100.0              |
|       | Total        | 60        | 100.0   | 100.0         |                    |

**Source: field data, October, 2014**

Regarding this question, the researcher decided to identify the kind of ICT tools used by the SMEs. The data collected showed that most of the ICT too used by these owners is the mobile phone. This represents 61.7% (37) respondents of the total sample size. 9 (15%) and 8 (13.3%) respondents use internet and electronic-money system respectively. 6(10%) respondents use the fixed line (telephone) and they represent the minority in this regard. The analysis indicates that though some of the SMEs do not have an idea of what ICT is, they do at least use one form of ICT. The inference drawn here is that at least each of the SMEs owners does employ one tool of ICT and this is a good sign for the market and an indication that the level of ICT usage in the New Juaben Municipality is good or high.

Table 4.4  What are the Reasons for not using ICT Infrastructure

|       |                          | Frequency | Percent | Valid Percent | Cumulative Percent |
|-------|--------------------------|-----------|---------|---------------|--------------------|
| Valid | High cost                | 33        | 55.0    | 55.0          | 55.0               |
|       | Ignorance about ICT benefits | 19    | 31.7    | 31.7          | 86.7               |
|       | Low technical know-how   | 8         | 13.3    | 13.3          | 100.0              |
|       | Total                    | 60        | 100.0   | 100.0         |                    |

**Source: field data, October, 2014**

Referring to the data captured above confirms that there are different reasons why some SMEs do not engage in ICT. Among the reasons include high cost involved in using ICT, ignorance about the benefits of ICT, Low knowledge about the usage. The frequency representations are 33, 19 and 8 respectively. Though from the previous analysis only 14 respondents said they don't use any tool, all the respondents do agree there are reasons with the highest saying high cost. The inference drawn here is that, there are reasons why some of the SME operators do not use ICT tools in their operations and the most pressing reason has to do the cost involved.

Table 4.5 Most common ICT tool(s) used by business

|       |              | Frequency | Percent | Valid Percent | Cumulative Percent |
|-------|--------------|-----------|---------|---------------|--------------------|
| Valid | Internet     | 7         | 15.0    | 15.0          | 15.0               |
|       | Mobile Phone | 39        | 61.7    | 61.7          | 76.7               |
|       | E-Money Sys. | 11        | 13.3    | 13.3          | 90.0               |
|       | Fixed line   | 3         | 10.0    | 10.0          | 100.0              |
|       | Total        | 60        | 100.0   | 100.0         |                    |

**Source: field data, October, 2014**

Asking about the preferred or most common ICT tool used, the data collected showed that, mobile phone is the most common. From the table above 39 (%) respondents indicated the most used is mobile phone, 11 (%) respondents said electronic money system, 7 (%) said internet usage and the least is the fixed line (telephone).The inference drawn is that SME operators prefer the use of mobile phone in their operations than fixed line as well as internet in communicating among business partners and clients. The analysis also shows that the second most used are the electronic money system, popularly known as money transfer. It shows again the level of usage of ICT among the SMEs is very high.

**Objective two:** To assess the level at which the adoption of Information And Communication Technology has led to the achievement of operational goals among the Small And Medium Scale Enterprises in the New Juaben Municipality.

Table 4.6  ICT Support System improved your operations

|       |       | Frequency | Percent | Valid Percent | Cumulative Percent |
|-------|-------|-----------|---------|---------------|--------------------|
| Valid | Yes   | 53        | 88.3    | 88.3          | 88.3               |
|       | No    | 7         | 11.7    | 11.7          | 100.0              |
|       | Total | 60        | 100.0   | 100.0         |                    |

**Source: field data, October, 2014**

With regards to ICT improving business operations of the SME owners, the data gathered proved positive. Majority of the respondents representing 88.3% said that ICT usage have improved their business operations. The reaming 7 (11.7%) respondents no ICT has not improved business operations. This number represents those who said they have no knowledge of ICT. The analysis here is that, majority of the SME operators in the new Juaben Municipality have seen improvements in their business as a results of ICT usage. This adds to the fact that they understand the need for ICT and explains the reason why majority use ICT to support their business.

Table 4.7 Role ICT plays in the survival of your business.

|       |                                         | Frequency | Percent | Valid Percent | Cumulative Percent |
|-------|-----------------------------------------|-----------|---------|---------------|--------------------|
| Valid | Tracked Bus. activities                 | 14        | 23.4    | 23.4          | 23.4               |
|       | Maintain close relationship with customers | 12     | 20.0    | 20.0          | 43.4               |
|       | Improved financial transaction of business. | 11   | 18.3    | 18.3          | 61.7               |

| | | Frequency | Percent | Valid Percent | Cumulative Percent |
|---|---|---|---|---|---|
| | Easy and fast communication | 23 | 38.3 | 38.3 | 100.0 |
| | Total | 60 | 100.0 | 100.0 | |

**Source: field data, October, 2014**

On the role ICT pays in their business operations, the Table above shows that the roles are many and very beneficial to the business. Majority of the respondents representing 23 (38.3%) said ICT helps to promote easy and fast communication between clients and business partners. 14 (23.4%) respondents said ICT helps track business activates even across borders.  12 and 11 respondents said that ICT helps to keep customers close and in-touch and also improves financial transactions respectively. The analysis revealed that the role ICT plays in SME business operations are numerous and such role helps move the business in diverse ways for the purpose of making them survive and grow as well.

**Table 4.8** ICT contribution to the survival of your business.

| | | Frequency | Percent | Valid Percent | Cumulative Percent |
|---|---|---|---|---|---|
| Valid | Lowered transaction cost | 9 | 15.0 | 15.0 | 15.0 |
| | lowered cost of communication | 37 | 61.7 | 61.7 | 76.7 |
| | Increased productivity | 8 | 13.3 | 13.3 | 90.0 |
| | Broadened market base | 6 | 10.0 | 10.0 | 100.0 |
| | Total | 60 | 100.0 | 100.0 | |

**Source: field data, October, 2014**

The researcher from the question above sought to know the contributions ICT offer to SMEs survival in the municipality. From the table above, majority representing 37 (61.7%) said ICT has lowered communication cost. 9 (15%) respondents said ICT has lowered transaction cost and the least representing 6 (10%) respondents that ICT has helped increased their market base. It can be said from the above table that ICT has contributed a lot to SMEs survival within the Municipality. It also shows that the SMEs operators appreciate the inclusion of ICT in their operation. This is a sign of good faith and excellent business operations among the SME in the near and far future.

**Table 4.9** Frequency in the usage of ICT tools in business operations

| ICT tool | Not very often | Not often | Neutral | Often | Very often |
|---|---|---|---|---|---|
| Electronic money system | 17 (28.3%) | 11(18.3%) | 13(21.7%) | | 19(31.7%) |
| Mobile phone | 0 | 5(8.3%) | 0 | 13(21.7%) | 42 (70%) |
| Fixed line (telephone) | 14(23.3%) | 39 (65%) | 0 | 3(5%) | 4(6.7) |
| Internet | 19 (31.7%) | 26 (43.3%) | 7(11.7%) | 0 | 8(13.3%) |

**Source: field data, October, 2014**

With regards to this question, it shows that the ICT tool used frequently in the SMEs operations is the mobile phone. This is followed by the electronic money system, internet and fixed lines. The operators explained that fixed line is gradually fading off as it is not handy and therefore does not support business operations outside the office. The reason for the mobile phone being the most patronized is because of its handy nature and also easy to apply coupled with internet facility within the phone. The analysis here is that the most preferred ICT tool in supporting business operations of the SMEs is the mobile phone.

**Objective three:** To find out the barriers/challenges of the usage of ICTs among SMEs in New Juaben Municipality.

All the respondents agreed, constituting 100% that they do face challenges on the usage of the ICT tools. The respondents explained that these challenges are of different nature as per the tool being used. Though majority agreed to understand the concepts of ICT usage in business operation, all the respondents here agree that the usage of ICT tools comes with a challenge and they do face them every single day of the business life. As a follow up question to the challenges the SMEs face in using ICT, the researcher sought to find out if these challenges do cost the operator money. From the data gathered, all respondents representing 100% indicated yes, Some of them said that break in communication network by the providers prolongs business and attract extra cost. Another challenge mentioned was software acquisition. Again, light out sometimes makes internet usage very worrying, hence causing the business to lose money.

**Table 4.12** Ranking the challenges in use of ICT

| Challenge | Strongly disagree | Disagree | Neutral | Agree | Strongly Agree |
|---|---|---|---|---|---|
| High cost of S.W applications | 0 | 0 | 0 | 7 (11.7%) | 53 (88.3%) |
| Lack of availability of internet and network infrastructure | 21 (35%) | 23 (38.3%) | 0 | 4 (6.7%) | 12 (20%) |
| Resistance of end-users to new technology | | 4 (6.7%) | 11 (18.3%) | 17 (28.3%) | 28 (46.7%) |
| Security and Trust Issues | | | | | |
| Lack of expertise in the areas of ICT in the industry | 3 (5%) | 0 | 12 (20%) | 8 (13.3%) | 37 (61.7%) |
| High Cost of ICT equipment and networks | 0 | 0 | 8 (13.3%) | 11 (18.3%) | 41 (68.4%) |

**Source: field data, October, 2014**

In ranking the challenges these SMEs face with respect to the use of ICT, the researcher found that the most pressing challenge is the cost of software needed to run the operations of the business. The next challenge hitting most is high cost of ICT equipment and networks known as the hardware. The SMEs explained that the cost of getting the communication gadgets is on a high side for them. Especially having a networked system in place to support sharing of information and data is very costly. On the other hand, computer systems have taken a trendy path making the cost also high and somehow expensive for some of the SMEs to afford. The third challenge they face has to do with Lack of expertise in the areas of ICT in the industry. Majority explained that the ICT experts focus greatly on the big players in the market and therefore design programs and software that support their operations leaving the small businesses. Even, these expertise come to SME market, they are not able to create for them programs that suit them. Some of the SME owners complained saying, the government has also not done much in the area of ICT for them.

With reference to the challenge, some of the SME operators admitted that resistance of end-users to new technology was another challenge they face. Most of their workers are semi-literate and illiterate thereby making the incorporation a bit difficult. What these workers understand is the use of mobile phone. Using networked system, computer, electronic payment system etc. is a major issue for some of the workers of the SME owner. However, they claimed that they are making effort to get these workers acclimatized to the usage of the ICT. Another challenge posed by the SME operators is issue about security and trust. Some said that they have been victims of lost information or otherwise money. Monies sent to partners went elsewhere due to wrong phone numbers. Messages sent either never went or got locked up. However, this challenge is not too great for them as they can always switch if need be especially with regards to lost or truncated information. Still another but the least challenge to them is the availability of network infrastructure. They said, they have been told such structure are available only that they find it a bit expensive and sometimes not worth the need to get one into the business.

**Discussion of findings**

**Objective one:** To ascertain level of ICT adoption by SMEs in their business operations

Regarding this objective, it was found out that a lot of the SMEs in New Juaben municipality do know and have an idea of ICT. It was also revealed that majority of the SME operators do use at least one ICT tool in supporting their operations within the New Juaben Municipality supporting the view that the number of SMEs currently using ICT in their business is very high. This shows that the SMEs owners understand the need to employ ICT in their operations. On the basis of the ICT tool these SMEs used in their business, the study revealed that each they use mobile phone, fixed line, and internet and electronic-money system. The study showed that at least each of the SMEs owners do employ one tool of ICT and this is a good sign for the market and an indication that the level of ICT usage in the New Juaben Municipality is good or high. The study also revealed that SMEs operators prefer the use of mobile phone in their operations than fixed line as well as internet in communicating among business partners and clients.

**Objective two:** To assess the extent to which the adoption of ICT has led to the achievement of operational goals.

Concerning this objective, it was found out that greater percentage i.e.(88.3%) of the SME operators in the New Juaben Municipality have seen improvements in their business as a results of ICT usage. The study revealed that ICT usage has helped the SMEs tracked business activities, Maintain close relationship with customers, improved financial transaction of business, easy and fast communication etc. It was evident that the role ICT plays in SMEs business operations are numerous and such role helps move the business in diverse ways for the purpose of making them survive and grow as well.

On the basis of the ICT to the survival of SMEs in the Municipality, the study showed that ICT has contributed a lot in that it has lowered transaction cost, lowered cost of communication, increased productivity, broadened market base etc. The study also showed that the SMEs operators appreciate the inclusion of ICT in their operations. The reason for the mobile phone being the most patronized is because of its handy nature and also easy to apply coupled with internet facility within the phone.

**Objective three:** To find out challenges in ICT usage among SMEs

On the basis of the challenges the SMEs face in using ICT tools, Though majority agreed to understand the concepts of ICT usage in business operation, all the respondents agreed that the usage of ICT tools comes with a challenge and they do face them every single day of the business life. Among them includes high cost of software, cost of ICT equipment and networks known as the hardware, lack of expertise in the areas of ICT in the SME industry, issues about security and trust, resistance of end-users to new technology.

The most pressing of the challenge as revealed by the analysis is cost of software needed to run the operations of the business. The SME operators explained that they know getting the software will help them a lot but the challenge is the cost as well as the financing of the software. It was also evident that another challenge they face is the non-availability of network infrastructure. The study revealed the SMEs do know that such structure are available only that they find it a bit expensive and sometimes not worth the need to get one into the business.

**Conclusion & Recommendations**

The research sought to assess the Role of Information And Communication Technology (ICT) in the Survival of Small and Medium Scale Enterprises (SMEs) : Evidence from Koforidua. From the research conducted, it is clear that there are quite worrying state of affairs in relation to the use of ICT in enhancing business growth strategies of SMEs within the Municipality. Generally, the respondents admitted that ICT can help improve the survival status of SMEs. However, the usage is limited to mobile phone leaving out chunk of the tools and their application or usage.

The role of ICT in improving business survival, delivery services and innovations in SME is ongoing but much needs o be done. ICT provides the bedrock on which SMEs can build their business information systems aimed at improving their business processes, customer relations and efficient delivery of goods and services to satisfy the needs of cherished customers. It is obviously clear from the findings that ICT infrastructure and innovations are insignificantly available in SMEs within the municipality. Reasons such as poor knowledge about ICT, lack of qualified personnel and high cost of software, high cost of hardware and its implementation were mainly cited as the challenge preventing the smooth implementation of ICT resources. Periodic training in the form of workshops and sensitization programs on the benefits and the use ICT resources in business growth strategies should be organized by National Board for Small-Scale Industries (NBSSI) for SME operators to create more awareness in order to enhance their preparedness to institute ICT programs aimed at improving their business operations. SMEs themselves should invest in educating their staff and management about ICT and its benefits.

As training cost is known to be one of the impediments towards the developing of ICT skills, in the end this will help these SMEs acquire these skills and use them in their work. There should be a comprehensive effort to institute ICT training programs at the various levels of the educational hierarchy. ICT should also be made compulsory at the basic and second cycle schools to train more students to increase ICT technical trainees and professionals to fill the ICT job market. This will also help as such individuals will grow to own SME business where such skills can be explored to their benefit.

SMEs and their owners should build a culture that is favourable to technology and innovative. SMEs owners should align technology to their business strategy and seek how technology can give them competitive advantage. In addition the government can enact a policy that will encourage the usage of ICT among these SMEs. Such policy may include registration of SMEs online, declaring of tax returns online, advertisement should include online etc. This policy when done will help increase ICT knowledge and usage within the municipality and the country at large. SME operators, can also outsource their ICT delivery systems by engaging ICT consultants in order to avoid the problem of funding relating to the setting up of their own ICT system which usually requires huge initial capital outlay. Again, the government, universities, and other institutions in the country can come out cheaper software and computers that are of durable quality to support these SMEs operations.

## BIODATA AND CONTACT ADDRESSES OF AUTHORS

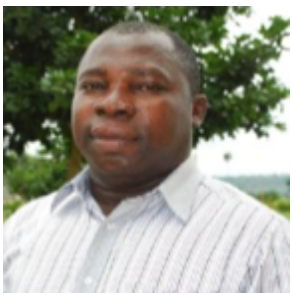

Martin Otu **OFFEI** is a lecturer at the Computer Science Department and currently the Director of the ICT Directorate at the Koforidua Polytechnic. He holds a Bsc and Msc degree in Computer Science.

Martin Otu   OFFEI
Koforidua Polytechnic, GHANA
E. Mail: martinoffei@yahoo.com

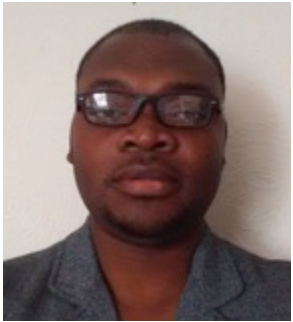

Kwaku **NUAMAH-GYAMBRAH** is a Lecturer in the Computer Science Department of Koforidua Polytechnic with over 10 years' experience in teaching and research. He holds an MSc in Software Technology with Network Management.

Kwaku NUAMAH-GYAMBRAH
Koforidua Polytechnic, GHANA
Email: kwaku.gyambrah@koforiduapoly.edu.gh

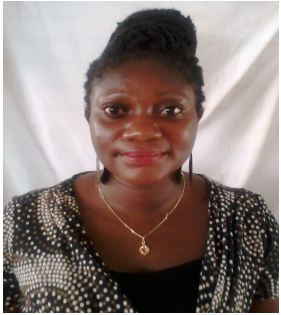

Florence Agyeiwaa is a Professional Teacher at Koforidua Senior High Technical School. She holds a 3-Year Teacher's Certificate 'A' (Agogo Presbyterian Teacher Training College), a graduate certificate in BSc. Hons Computer Science from the All Nations University College and a Master's Degree in Business Information Technology from the Kwame Nkrumah University of Science and Technology (KNUST) Business School.

Florence **AGYEIWAA**
All Nations University College
Koforidua Ghana

**References**

[1.] Acolatse, S. (2012). Challenges & opportunities for SME development in Ghana, S.A.

[2.] Acquah (2012), Saturday, Banks urged to treat Ghanaian SMEs special, October 27, 2012

[3.] Aryeetey, E. (2005). Small Enterprise Credit In West Africa. London, The British Council.

[4.] Attom, M (2008) "Information Systems Strategy and Knowledge-based SMEs in the Australian Biotechnology Industry: Does IS need to reorient its Thinking?" 15th Australian Conference on Information Systems, Hobart.

[5.] Barman, J (2010). Toward Strategic Use of IT in SMEs: "A Developing Country Perspective", Information Management & Computer Security Vol. 11, No. 5, pp. 230-237

[6.] Brady et al., (2002). Performance-only measurement of service quality: a replication and extension. A Journal of Business Research, Vol 55, 17 – 31.

[7.] Braun (2004) The New World of Microenterprise Finance: Building Healthy Financial Institutions for the Poor, West Hartford, CN: Kumarian Press.

[8.] Cela, J (2005). SME development in Indonesia with reference to networking, innovativeness, market expansion and government policy. Readworthy Publications, Ltd, New Delhi

[9.] Computer Security Institute. 2003. CSI/FBI Computer Crime and Security Survey. Computer Security Institute: San Francisco, CA

[10.]   Curran, J. (1997), 'The role of the small firm in the UK economy: hot stereotypes and cool assessments', Small Business Research Trust, May, Milton Keynes: Open University.

[11.]   Esselaar et al, (2007).Supporting the E-business Readiness of Small and Medium-Sized Enterprises: Approaches and Metrics", Internet Research Vol. 12, No. 2, pp139-195.

[12.]   Ghana Banking Survey (2004). Price Waterhouse Coopers reports

[13.]    Ghana News Agency. (2006). SMEs should be given necessary support.

[14.]    Joel Conman and Robert N. Lussien (2006). Understanding the Successful Adoption and Use of IS/IT in SMEs: an Explanation from Portuguese Manufacturing Industries", Information Systems Journal, Vol. 12, No. 2, pp 121-152.

[15.]    Johnston and Lawrence, 1998; Kahn, 1996, 2001).SMEs and Barriers to skills development: A Scottish Perspective", Journal of European Industrial Training, Vol. 24, No. 1, pp. 5-11.

[16.]    Laudon, K.C. & Laudon, J.P. (2010). Management Information System, New Jersey: Pearson Education Inc.

[17.]    McMullan, J., & Perrier, D. (2003). Technologies of crime: The cyber-attacks on electronic gambling machines. Canadian Journal of Criminology and Criminal Justice, 45(2), 159-186.

[18.]    Ojo (2009).Development of Small and Medium Scale Enterprises: The role of Government and other Financial Institutions. Arabian Journal of Business and Management Review (OMAN Chapter) Vol. 1, No.7; February 2012

[19.]    Porter M (2005): How information gives you competitive advantage Harvard Business Review Vol 63 Issue 4 Jun/July 1985 pp 149-160

[20.]    Porter M, Millar VE (2005): How information gives you competitive advantage Harvard Business Review Vol 63 Issue 4 Jun/July 1985 pp 149-160

[21.]    Porter, O and Millar, D. (2005). The Importance of ICT: An Empirical Study in Swiss SMEs" in 19th Bled Conference eValues, Bled, Slovenia, June 5-7.

[22.]    Ricky-Okine, Twum and Owusu, (2014).A qualitative approach to examining the challenges of Ghanaian small and medium scale enterprises. (SMEs):

[23.] Rouf, K. A. (2012). The advantages of micro-credit lending programs and the human capabilities approach for women's poverty reduction and increased human rights in Bangladesh. International Journal of Research Studies in Management, 1(2).